\documentclass[conference]{IEEEtran}
\usepackage{nomencl}

\usepackage{cite}

\usepackage{multirow}

\ifCLASSINFOpdf
   \usepackage[pdftex]{graphicx}
\else
\fi
\usepackage{amsmath}
\usepackage{array}
\usepackage{url}

\usepackage[fleqn]{nccmath}

\begin{document}
%
\title{Assessment of the Value of Frequency Response \\Times in Power Systems}

\author{\IEEEauthorblockN{Yifu Ding}
\IEEEauthorblockA{Energy and Power Group\\
Department of Engineering Science\\
University of Oxford, UK\\
Email: yifu.ding@eng.ox.ac.uk}
\and
\IEEEauthorblockN{Roberto Moreira}
\IEEEauthorblockA{Department of Electrical and\\Electronics Engineering\\
Imperial College London, UK\\
Email: roberto.m@imperial.ac.uk}
\and
\IEEEauthorblockN{Dagoberto Cedillos}
\IEEEauthorblockA{Open Energi\\
239 Old Marylebone Road, London, UK\\
Email: dagocedillos@gmail.com}}

\maketitle

\begin{abstract}
Given the increasing penetration in renewable generation, the UK power system is experiencing a decline in system inertia and an increase in frequency response (FR) requirements. Faster FR products are a mitigating solution that can cost-effectively meet the system balancing requirements. Thus, this paper proposes a mixed integer linear programming (MILP) unit commitment model which can simultaneously schedule inertial response, mandatory FR, as well as a sub-second FR product - enhanced frequency response (EFR). The model quantifies the value of providing faster reacting FR products in comparison with other response times from typical FR products. The performance and value of EFR are determined in a series of future energy scenarios with respect to the UK market and system conditions. \\

\end{abstract}

\IEEEpeerreviewmaketitle

\section{Nomenclature}

\begin{table}[!h]
\label{table_example}
\normalsize
\renewcommand\arraystretch{0.95}
\begin{tabular}{lp{2.7in}}
\textit{ A.}& \textit{Set and indices}\\
$t$& Set of time periods (h)\\
$g$& Set of thermal generation unit groups\\
$s$& Set of pumped-hydro storage (PHS) \\
&\\
\textit{ B.}& \textit{Constants}\\
$E^s_{max}$&Maximum energy level of unit s (MWh)\\
$E^s_{min}$&Minimum energy level of unit s (MWh)\\
$\eta_{s}$&Round-trip efficiency of the unit s\\
$P^{s}_{c max}$&Maximum charging power of unit s (MW)\\
$P^{s}_{d max}$&Maximum discharging power of unit s (MW)\\
$C_{g}$& Nameplate capacity of a generation unit g (MW)\\
$c^{m}_{g}$&Marginal cost of a unit in group g ($\pounds$/MWh)\\
$c^{nl}_{g}$&No-load cost of a unit in group g  ($\pounds$/h)\\
$c^{st}_{g}$&Start-up cost of a unit in group g  ($\pounds$)\\
$F^{pr max}_{g}$&Maximum primary frequency response (PFR) provision of a unit in group g  (MW)\\
$F^{se min}_{g}$&Maximum secondary frequency response (SFR) provision of a unit in group g (MW)\\
$F^{max}_{g}$&Maximum FR provision of unit s (MW)\\
$N_{g}$&The number of units in group g\\
$\rho^{th}_{g}$&PFR slope coefficient\\
$P^{msg}_{g}$&Minimum stable generation (MSG) level for a unit in group g (MW)\\
\end{tabular}
\end{table}

\begin{table}[!h]
\label{table_example}
\normalsize
\renewcommand\arraystretch{0.95}
\begin{tabular}{lp{2.7in}}
$t^{start}_{g}$&Start-up time for a unit in group g (h)\\
$t^{shut}_{g}$&Shut-down time for a unit in group g (h)\\
$V^{up}_{g}$&Maximum ramp-up rate of a unit in group g (MW/h)\\
$V^{dn}_{g}$&Maximum ramp-down rate of a unit in group g (MW/h)\\
$w_{t}$&Wind power output in period t (MW)\\
$I_{t}$&Imported/exported power from the interconnection in period t (MW)\\
$d_{t}$&Total system demand in period t (MW)\\
$O_{t}$&Grid-connected solar power output in t (MW)\\
$f_{o}$&Nominal system frequency (Hz)\\
$f_{nadir}$&Maximum frequency deviation at nadir (Hz)\\
$f_{ss}$&Maximum frequency deviation at quasi-steady-state(QSS) (Hz)\\
$h_{g}$&Inertia constant of group g (s)\\
$h_{l}$&Inertia constant of load l (s)\\
$P^{max}_{l}$& Maximum infeed loss (MW)\\
$D$&Load damping rate (1/Hz)\\
$T_{p}$&Response time of PFR (s)\\
$T_{e}$&Response time of EFR (s)\\
$E$&Total amount of EFR (MW)\\
&\\
\textit{ C.}& \textit{Variables}\\
$P^{th}_{t,g}$&PFR provision by group g in period t (MW)\\
$P^s_{t}$&PFR provision by unit s in period t (MW)\\
$S^{th}_{t,g}$&SFR provision by group g in period t (MW)\\
$S^s_{t}$&SFR provision by unit s in period t (MW)\\
$s^{on}_{t,g}$&Number of generators turned on in group g in period t\\
$s^{off}_{t,g}$&Number of generators turned off in group g in period t\\
$u^{th}_{t,g}$&Number of online generators in group g in period t\\
$P^{Gen}_{t,g}$&Power output of groups g in period t (MW)\\
$P^{c}_{t}$&Charging power of unit s in period t (MW)\\
$P^{d}_{t}$&Discharging power of unit s in period t (MW)\\
$E^{s}_{t}$&Instantaneous energy in unit s in period t (MWh)\\
$P^{req}_{t}$&PFR requirement in period t (MW)\\
$S^{req}_{t}$&SFR requirement in period t (MW)\\
$H_{t}$&System inertia in period t (MVAs)\\
\end{tabular}
\end{table}

\section{Introduction}
To mitigate the aggravation of climate change and accommodate the challenges with the decarbonization of other energy sectors e.g. heat and transport -, the UK electricity system is facing unprecedented changes. The increasing penetration of low carbon generation combined with the decommission of conventional generation plants is expected to significantly undermine power system inertia, and thus system stability. As a significant proportion of the UK generation mix will be based on volatile and uncertain low carbon generation, the need for system frequency regulation services will also increase dramatically. Moreover, insufficient levels of system inertia€" due to the deficit in large rotating masses - can lead to significant post-contingency system frequency deviations, and thus further aggravate the need for frequency regulation services. This leads to lower asset utilization factors, inefficiency costs associated with partly-loaded generation plants, curtailment of low carbon generation and overall to higher operational costs. 

In this setting, frequency regulation products such as the case of Enhanced Frequency Response (EFR) in UK, characterized by faster reaction times (i.e $\leq$1 s) than other conventional FR products (i.e. 10s) allow deviations in system frequency to be swiftly mitigated \cite{NationalGrid.2017}. Moreover, the improved efficiency of EFR mitigates frequency deviations, allows for lower committed volumes of FR products and therefore lower system balancing costs. Currently, a large range of technologies are capable of delivering EFR services, including some types of energy storage systems (e.g. batteries, flywheels and supercapacitors) and advanced demand response control \cite{Cooke2017}. This paper focuses on evaluating their economic benefits of EFR and identifies the influential factors of its performance. It helps stakeholders owning the aforementioned assets find investment opportunities in the present and future balancing markets. 

\subsection{Related work}

Previous works estimate FR requirements by adopting different methodologies, from pre-defined, generation-based requirements to inertia-dependent ones. In order to assess the economic benefits of dynamic demand in providing frequency regulation, \cite{Aunedi2013} constructed a unit commitment (UC) model with per-defined frequency regulation and reserve requirements. Similarly, \cite{Vijay2017} developed a comprehensive UC model for the UK power system and used historical reserve requirements to estimate future electricity and reserve prices. System inertia can capture transient frequency evolution dynamics to accurately determine hourly-based FR requirements. \cite {Chavez2014a} uses the system inertia to estimate PFR adequacy. Recent works \cite{Teng2017} successfully conducted the stochastic scheduling for generation dispatch and ancillary services with inertia-dependent constraints. Regarding the multi-speed FRs, \cite{Trovato2018} formulates the model that can simultaneously allocate inertia responses (IR), PR and EFR, and evaluates the EFR effectiveness. Considering different inertia conditions, \cite{Greve2017} proposed a utility function for system operator (SO) to choose multi-speed FR products.

\subsection{Contribution}

Compared with the previous work in the literature, the key contributions are highlighted as follows:

1) The work considers a full set of FR services in the UK frequency regulatory framework including inertia response, primary frequency response (PFR) and secondary frequency response (SFR). We evaluate the offset effect of EFR against those two mandatory FRs as an intuitive indicator for the market. 

2) The temporal variation in system inertia and FR requirements are acquired, which thus identifies the influential factors of hourly EFR performance. 

\section{Unit commitment model}

The proposed mixed integer linear programming (MILP) model aims to schedule multi-speed frequency responses and estimate costs in current UK balancing market framework without adopting the computationally expensive algorithm. This work uses the deterministic formulation, while the stochastic scheduling algorithm such as \cite{Teng2016} performs better in terms of reducing wind curtailment. For simplicity, the model aggregates power plants with the same technology into one group. The model focuses on the low frequency responses which address the in-feed loss, and other ancillary services in the balancing market such as reserve is out of the research scope. We assume that there exists the utility-scale energy storages to provide EFR. 

\subsection{Unit Commitment Model with Inertial Constraints}

The objective function (1) describes the (annual) system operation cost (i.e. the sum of the start-up, no-load and marginal costs of conventional generation units. 

\begin{fleqn}
\begin{equation} 
\label{eq:1}
\scalebox{0.9} {$
 f=min {\sum_{g=1}^{G}\sum_{t=1}^{T}(c^{st}_t\cdot{s^{on}_{t,g}}+c^{nl}_i\cdot{u^{th}_{t,g}}+c^{m}_i\cdot{P^{Gen}_{t,g}})} $}
 \end{equation}
\end{fleqn}

The demand-supply balance is ensured in (2). In (3)-(7) the model captures the binary operational characteristic of conventional units, i.e. online / offline times€" with exception of nuclear power plants (i.e. $G_n$) as (3); (4) defines minimum stable generation (MSG) and maximum generation levels. (5) calculates the number of shut-down and start-up in each generator group. (6) defines the minimum start-up and shut-down time. (7) defines the maximum ramp-up and ramp-down rates. The model also includes energy storage units technical limitations, such as maximum and minimum energy levels (8), maximum charge/discharge capacities (9), unit€™s energy balance (10) and conservation of energy (11). 
The initial state of scheduling is pre-defined (i.e. $u^{th}_{t0,g}$, $P^{Gen}_{t0,g}$). For all the constraints if not specified, $\forall$g$\in$G and $\forall$t$\in$T.

\begin{fleqn}
\begin{equation}
\scalebox{0.9}{$
\sum_{g=1}^{G}P^{Gen}_{t,g}+P^{d}_{t}+I_{t}+\omega _{t}+O_{t}={d_{t}+P^{c}_{t}}$}
\end{equation}
\begin{equation}
\scalebox{0.9}{$
  \begin{cases}
   u^{th}_{t,g}=N_{g} &\quad  g\in G_{n}\\
   u^{th}_{t,g} \leq N_{g}& \quad   g\in G_{t}
  \end{cases} $}
  \end{equation}
  \begin{equation}
 \scalebox{0.9}{$
{u^{th}_{t,g}}\cdot{P^{msg}_{g}}\leq{P^{Gen}_{t,g}}\leq{u^{th}_{t,g}}\cdot{C_{g}} $}
\end{equation}
\begin{equation}
\scalebox{0.9}{$
  \begin{cases}
  s^{on}_{t,g}=u^{th}_{t,g}-u^{th}_{t-1,g}\\ 
  s^{off}_{t,g}=u^{th}_{t-1,g}-u^{th}_{t,g}
  \end{cases}$}
  \end{equation}
  \begin{equation}
 \scalebox{0.9}{$
  \begin{cases}
   u^{th}_{t,g}\geq \sum_{\tau=t-{t_{g}^{start}+1}}^{t-1}{s_{t,g}^{on}}\\ 
   u^{th}_{t,g}\leq {G_{g}}- \sum_{\tau=t-{t_{g}^{start}+1}}^{t-1}{s_{t,g}^{on}}
  \end{cases} $}
  \end{equation}
  \begin{equation}
  \scalebox{0.9}{$
   \begin{cases}
   P^{Gen}_{t,g}-P^{Gen}_{t-1,g}\leq {u_{t,g}^{th}\cdot{V_{g}^{up}}}\\ 
   P^{Gen}_{t-1,g}-P^{Gen}_{t,g}\leq {u_{t,g}^{th}\cdot{V_{g}^{dn}}}
  \end{cases}$}
  \end{equation}
  \begin{equation}
  \scalebox{0.9}{${E^{s}_{min}}\leq{E^{s}_{t}}\leq{E^{s}_{max}}$}
  \end{equation}
  \begin{equation}
  \scalebox{0.9}{$ \begin{cases}
   P^{c}_{t}\leq {P_{c max}^{s}}\\ 
   P^{d}_{t}\leq {P_{d max}^{s}}
  \end{cases}$}
  \end{equation}
\end{fleqn}

\begin{fleqn}
\begin{equation}
\scalebox{0.9}{${E^{s}_{t}}={E^{s}_{t-1}}+({\eta_{s}}\cdot{P^{c}_{t}}-\frac{P^{d}_{t}}{\eta_{s}})$}
\end{equation}
  \begin{equation}
  \scalebox{0.9}{${E^{s}_{0}}={E^{s}_{T}}$}
  \end{equation}
\end{fleqn}

We use the inertia-dependent constraints to estimate FR requirements. The aggregated system inertia given in (12) is the sum of inertia from all online thermal generators and responsive loads (i.e. rotating motors).
\begin{fleqn}
\begin{equation}
\scalebox{0.9}{${{H_{t}}={\sum_{g=1}^{G}{{c_g}\times{h_g}\times{u^{th}_{t,g}+{d_t}\times {h_l}}}}}$}
\end{equation}
\end{fleqn}
The timescale of three FR products, including PFR (10-30s), SFR (30s-30min), and EFR ($\leq$1s), is shown in Fig. 1. 
 \begin{figure}[h!]
  \begin{center}
  \vspace*{-0.1in}
    \includegraphics[width=3.0in]{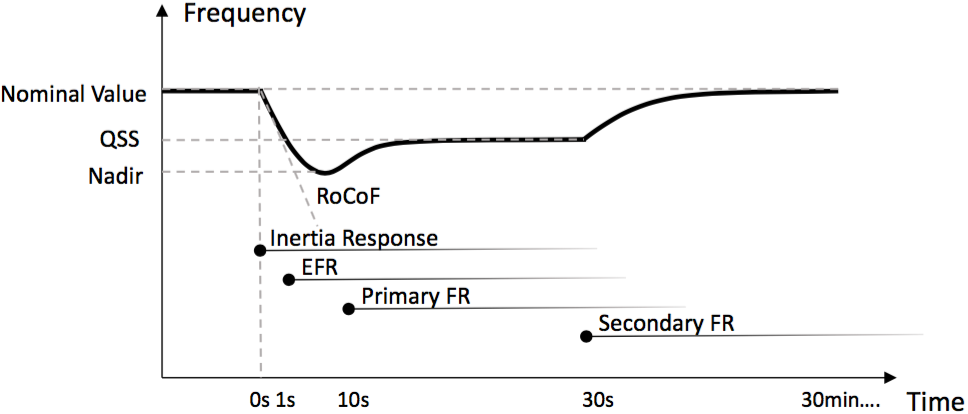}
  \vspace*{-0.1in}
  \end{center}
  \caption{The frequency evolution in the power system after a contingency}
\end{figure}

The swing equation (13) models the relationship between system in-feed loss and frequency evolution. The model adopts $N-1$ reliability criterion. The in-feed loss level for regulating FR requirement is set to be 1320 MW \cite{Cheng2016}. This work does not consider the dead band happening in the real FR provision process. 

\begin{fleqn}
\begin{equation}
\scalebox{0.9}{$
{2{H_{t}}\frac{\partial \delta f(t)}{\partial t}+D\cdot{d_t}\cdot{\delta f(t)}= \delta {P_L} - R}$}
\end{equation}
\begin{equation}
\scalebox{0.9}{$R=
\begin{cases}
 E /{T_e} &\quad 0\leq t < {T_e} \\ 
 E+ {P^{req}_t}/ {T_p} &\quad  {T_p}\leq t \leq{T_e} \\
{P^{req}_t}  &\quad t > {T_e}
 \end{cases}
 $}
 \end{equation}
\end{fleqn}

Due to the in-feed power loss, the system frequency will drop from the nominal value (i.e. 50Hz). The inertia response happens from the onset of the loss. PFR are then triggered to arrest the frequency from falling until it reaches the nadir, where the rate of change of frequency (RoCoF) reaches zero (i.e. $(\partial f(t))/\partial t=0$) and the frequency stops falling. Based on (13), frequency responses ensuring that frequency reaches the nadir can be formulated as (15). 

At the nadir, the frequency should not exceed the regulated value in UK Grid Code (i.e. 49.2Hz). For simplicity, we omit the load damping term (i.e. $D\cdot{d_t}\cdot{\delta f(t)}$) in (13), integrate the equation regarding time, and substitute $R$ in (14) to obtain (16). It should be noted that (16) as a bilinear constraint needs to be linearized as Fig. 2. From the 30s after the contingency, SFR is delivered to replace PFR. It aims to recover the system frequency to the Quasi-steady-state (QSS) level, where RoCoF is zero. The total requirements (i.e. SFR and EFR) satisfying the condition can be estimated as (17). 

 \begin{figure}[h]
  \begin{center}
   \vspace*{-0.1in}
    \includegraphics[width=2.2in]{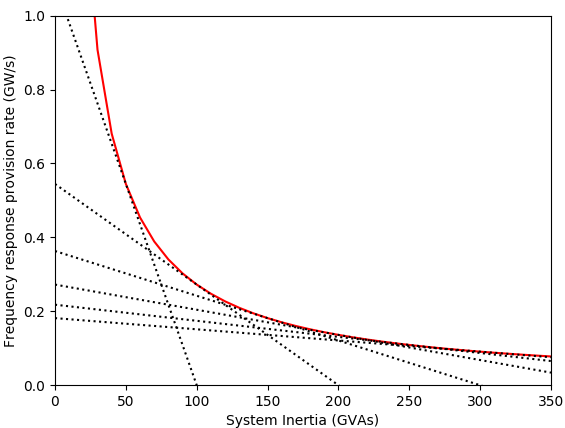}
   \vspace*{-0.1in}
  \end{center}
  \caption{Bilinear constraint linearization}
\end{figure}

   \vspace*{-0.2in}
\begin{fleqn}
\begin{equation}
\scalebox{0.9}{$
{P_{t}^{req}} + E \leq {{P_{l}^{max}}-D\cdot {d_t}\cdot{f_{max}}}$}
\end{equation}
\begin{equation}
\scalebox{0.9}{$
\frac {H_t}{f_o} \cdot (E + \frac {P_{t}^{req}} {T_p}) \leq \frac {{P_{t}^{req}}\cdot {P_{t}^{req}}}{4(f_o-f_{max})} $} 
\end{equation}
\begin{equation}
\scalebox{0.9}{${S_{t}^{req}}+E \leq {{P_{l}^{max}}-D\cdot {d_t}\cdot{f_{ss}}}$}
\end{equation}
\end{fleqn}

For a thermal power plant, FR provision are achieved by the governor control system. According to \cite{Erinmez1999}, the provision characteristics can be captured by (18) and (19). (20) defines the maximum PFR and SFR provision levels from storages. (20) and (21) define that FR provision must meet requirements.

\begin{fleqn}
\begin{equation}
\scalebox{0.9}{$
{P_{t,g}^{th}}\leq{min({F^{pr max}_{g}},\rho^{th}_{g}(u^{th}_{t,g}\times{C_g}-{P_{Gen}^{t,g}}))}$}
\end{equation}
\begin{equation}
\scalebox{0.9}{$
{S_{t,g}^{th}}\leq{min({F^{se max}_{g}}, u^{th}_{t,g}\times{C_g}-{P_{Gen}^{t,g}})}$}
\end{equation}
\begin{equation}
\scalebox{0.9}{$
{P_{t,g}^{s}},{S_{t,g}^{s}}\leq{min({F^{max}_{s}}, {P_{d max}^{s}} - {P_{d}^{t}})}$}
\end{equation}
\begin{equation}
\scalebox{0.9}{$
{P_{t}^{req}}\leq{\sum_{g=G_t}^{G}{P^{th}_{t,g}}+{P^{s}_{t,g}}}$}
\end{equation}
\begin{equation}
\scalebox{0.9}{$
{S_{t}^{req}}\leq{\sum_{g=G_t}^{G}{S^{th}_{t,g}}+{S^{s}_{t,g}}}$}
\end{equation}
\end{fleqn}

%
%


\section{Model results}

The model was performed on both baseline and future scenarios. The baseline case envisages current UK power system, while future scenarios (i.e. 'Steady State' and 'Two Degrees') are 2020 and 2025 energy system projections from National Grid Future Energy Scenario (FES) \cite{NationalGrid.2017a}. 

\vspace*{-0.1in}

\begin{table}[!h]
\caption{Thermal power plants parameters in the baseline case}
\label{table_example}
\centering
\begin{tabular}{lcccc}
\hline
&\textbf{Nuclear} & \textbf{Coal}& \textbf{CCGT} & \textbf{OCGT}\\
\hline
Nameplate capacity (MW) & 1800 & 500 & 500 & 200\\
Start-up cost (\pounds) & 50548 & 21001& 12564 & 0\\
Marginal cost (\pounds/MWh)& 7.1&19.8&18.93&39.54\\
No-load cost (\pounds/h) &0 &2071 & 2476 & 4809\\
MSG (MW) &1800 &200 & 200 & N/A\\
Start-up time (h)& 0&4 & 2 & 0 \\
Shut-down time (h) & 0 & 4 & 2 & 0\\
Governor slope & 0 & 0.3 & 0.4 & 0.6 \\
Inertia constant (s) & 4 & 6 & 6 & 6 \\
Ramp-down rate (MW/h) & 0 & 240 & 360 & 200 \\
Ramp-up rate (MW/h) & 0 & 200 & 360 & 200 \\

\hline
\end{tabular}
\end{table}

The model inputs include hourly demand and interconnection flow\cite{ElexonLimited2017}, hourly renewable generation capacity factors including solar PV, onshore and offshore wind farm\cite{Pfenninger2016}, technical parameters of thermal power plants \cite{SingleElectricityMarketCommittee2011} and PHS parameters. Future cycling costs of thermal power plants were calculated according to their fuel consumptions, annual average carbon and wholesale fuel prices for the power generation sector from \cite{NationalGrid.2017a} and \cite{Ofgem2016}. The system load including rotating motors is assumed to have the inertia constant 1s.

Results demonstrate increasing FR requirements as well as their costs in the future. Table II shows system inertia, FR requirements and operation costs in five scenarios. PFR requirements increase significantly, especially under the high renewable penetration (i.e. 'Two Degrees' scenario). SFR requirement keeps the same level as it aims to the post-fault energy restoration. Its volume is decided by in-feed loss, QSS and system load\cite{Vogler-Finck2015}, which are constant in this work. The balancing cost comes from the fuel and carbon costs when extra part-loaded units are operated to provide FR.

\vspace*{-0.1in}

\begin{table}[!h]
\scriptsize 
\caption{The summary of results in five scenarios}
\label{table_example}
\centering
\begin{tabular}{p{2.45cm}p{0.7cm}p{0.7cm}p{0.7cm}p{0.7cm}p{0.7cm}}
\hline
&\textbf{Base-} & \textbf{2020}& \textbf{2025} & \textbf{2020} & \textbf{2025}\\
&\textbf{line} & \textbf{Steady}& \textbf{Steady} & \textbf{Two} & \textbf{Two}\\
&& \textbf{State}& \textbf{State} & \textbf{Degrees} & \textbf{Degrees}\\
\hline
\textbf{Renewable pct. (\%)} & \textbf{25.2} & \textbf{44.4} & \textbf{47.2} & \textbf{47.0} & \textbf{58.5}\\
\hline
Max sys. inertia (GVAs) & 340 & 305 & 295 & 292 & 274\\
Min sys. inertia (GVAs) & 128 & 125 & 125 & 120 & 116\\
\hline
\textbf{Avg. sys. inertia(GVAs)}&\textbf{198} & \textbf{125} & \textbf{125} & \textbf{120} & \textbf{116}\\
\hline
PFR requirement (MW) &1366 & 1650 & 1726 & 1774 & 2049\\
SFR requirement (MW) & 1159 & 1159 & 1159 & 1159 & 1159\\
\hline
\textbf{Total FR (MW)} & \textbf{2525} & \textbf{2809} & \textbf{2885} & \textbf{2933} & \textbf{3208}\\
\hline
Energy cost (mil. \pounds) & 4340 & 4576 & 4694 & 3899 & 1929\\
Balancing cost (mil. \pounds) & 77 & 243 & 415 & 347 & 907\\
\hline
\textbf{Total costs (mil. \pounds) } & \textbf{4417} & \textbf{4819} & \textbf{5109} & \textbf{4246} & \textbf{2836}\\
\hline
\end{tabular}
\end{table}

\vspace*{-0.1in}
\subsection{Overall and hourly performance of EFR}

To examine the overall EFR performance, we carried out the simulation with the EFR deployment. Fig. 3 shows seasonal FR requirements with/without deploying 100MW EFR in the  baseline case. As a result, EFR can offset a large volume of PFR, but cannot offset SFR. The result highlights that EFR can offset a highest volume of PFR in Summer (i.e 700 MW on average), when both system inertia and demand are low.

 \begin{figure}[!h]
  \begin{center}
    \includegraphics[width=3.5in]{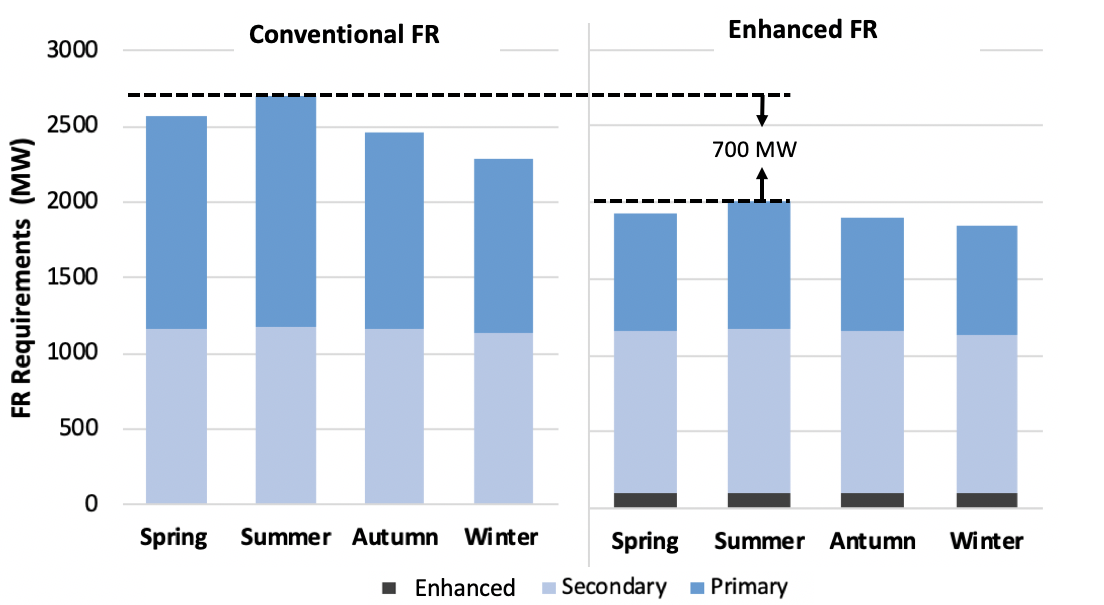}
  \end{center}
  \vspace*{-0.1in}
  \caption{Seasonal frequency response requirements in the baseline case before (left) and after (right) deploying 100 MW EFR}
\end{figure}

To further investigate hourly performance of EFR and its influential factors, According to the results of 8760 hours in the baseline case, we first calculated annual system inertia distribution (i.e. x axis) and demand distribution (i.e. y axis), then we plotted and color-coded the offset PFR by 1 MW EFR regarding corresponding inertia and demand in every time interval. Two critical conclusions can be drawn from the chart (Fig. 4). First, system demand and inertia are non-independent and have strong positive correlation. Second, the hourly performance of EFR mainly depends on the system inertia and demand. Considering facts above, the EFR effectiveness is a bivariate distribution regarding system inertia,  demand and their covariance.

  \vspace*{-0.1in}
  
 \begin{figure}[!h]
  \begin{center}
    \includegraphics[width=3in]{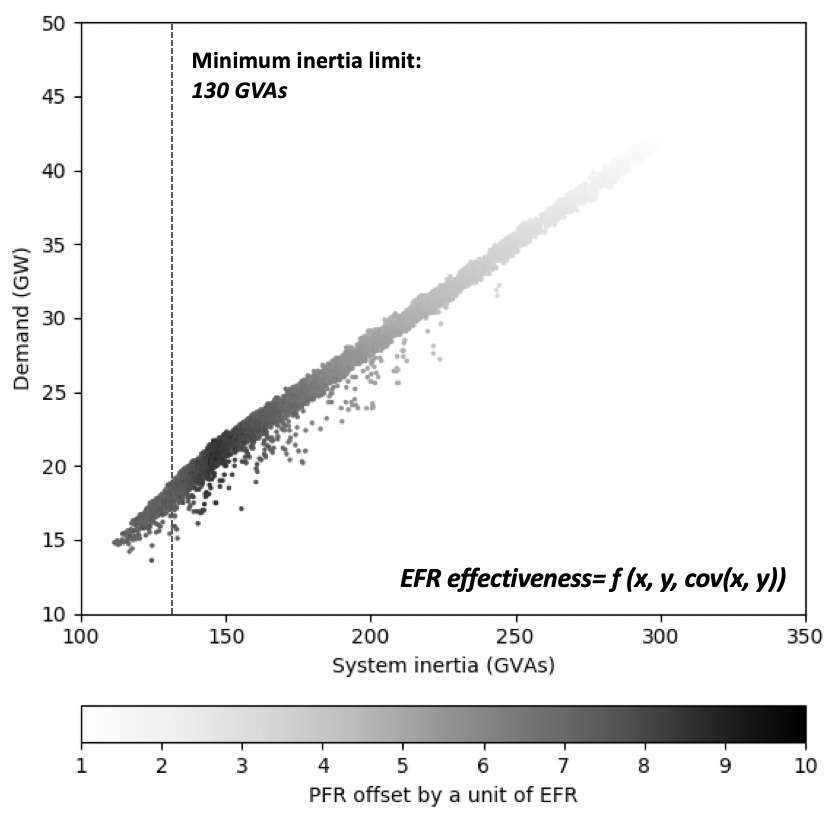}
  \end{center}
  \caption{Hourly EFR effectiveness regarding annual system inertia and demand distribution}
\end{figure}

  \vspace*{-0.1in}
\subsection{Economic benefits of EFR}


\begin{figure}[!b]
\begin{center}
\vspace*{-0.1in}
\includegraphics[width=2.7in]{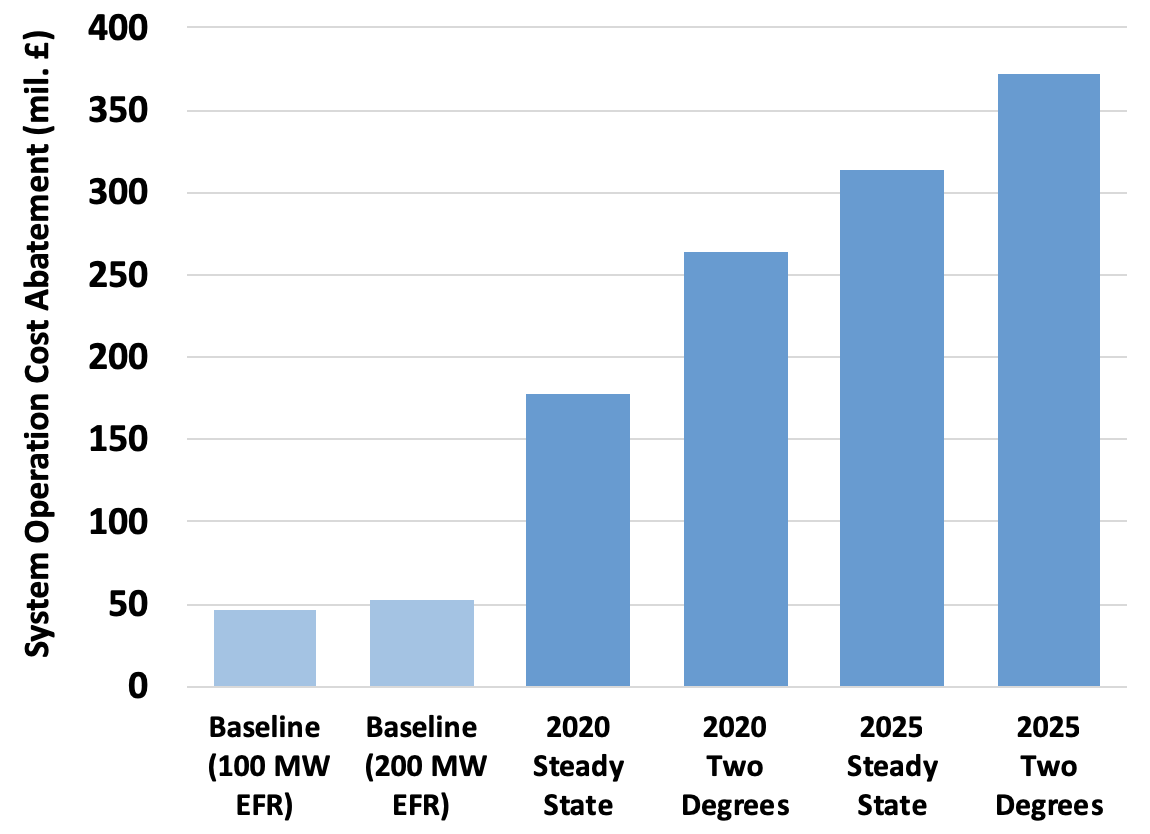}
\end{center}
\vspace*{-0.1in}
\caption{System operation cost abatements after deploying the EFR under different scenarios (ranking according to renewable penetration pct. from left to right)}
\end{figure}

The economic benefits of EFR are calculated as the abatement of system operation costs, compared with the benchmark without EFR services. We simulated baseline and four future scenarios with 200MW EFR deployment. The results are summarized in Fig. 5. First, results clearly show the economic benefit of EFR increases as the renewable penetration in the power system increases. Second, it demonstrates that EFR can reach the saturation level - there is no significant improvement in saving when EFR deployment increases from 100 to 200 MW in the baseline case.

\begin{figure}[!h]
\begin{center}
\vspace*{-0.1in}
\includegraphics[width=3.5in]{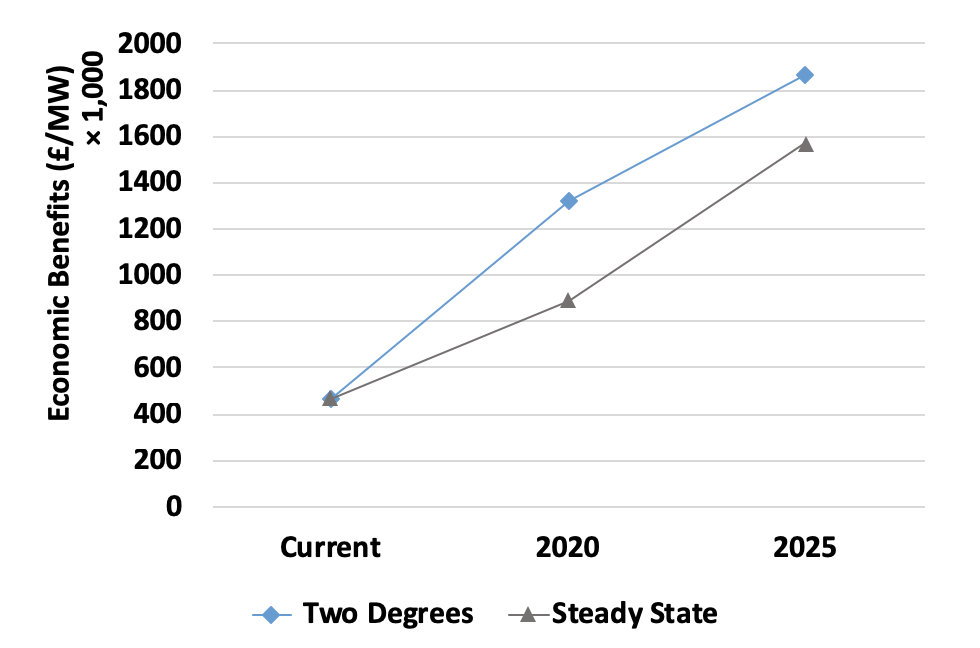}
\end{center}
\vspace*{-0.1in}
\caption{The economic benefits of EFR (thousand \pounds /MW) in 'Steady State' and 'Two Degrees' scenarios}
\end{figure}

Fig. 6 shows the value of per MW EFR for each scenarios. In scenarios with higher penetration of low carbon generation (i.e. 'Two Degrees'), the value of EFR is expected to increase by fourfold. However, the increasing rate in 'Two Degrees' scenario during 2020 - 2025 tends to less than the rate in 'Steady State'. The reason is that EFR relaxes FR requirements considerably so that the number of online part-loaded thermal plants decreases. That will further undermine system inertia and increase FR requirement conversely.
  
\section{Conclusion}
This paper demonstrates, through a MILP UC model with inertia-dependent constraints, the economic benefits of faster reaction times for FR products. In a series of case studies, with respect to different penetration scenarios of low carbon generation, it has been shown that faster FR products - such as EFR - can more efficiently mitigate deviations in system frequency and therefore leading to lower requirements for FR (i.e. lower MW available).


Moreover, this paper also analyses the future value of EFR across future low carbon generation scenarios. As a large proportion of low carbon generation is expected to replace conventional, high inertia-based thermal generation, the overall value of EFR is expected to increase. Indeed, the value of EFR tends to increase in the inverse proportion to instantaneous system inertia and demand.

To conclude, this research work provides an underlying analysis of the value of faster reacting FR products. This work provides an evidence-based reference for SO and other key stakeholders to understand and plan for adequate balancing market mechanisms to fully support the integration of low carbon generation in the UK. 

\section{Acknowledge}
The authors wish to thank Energy Future Lab at Imperial College London for the dissertation award and Energy and Power group at the University of Oxford for travel grants.




\bibliographystyle{IEEEtran}
\bibliography{/Users/apple/Desktop/paper/bibtex/IEEEconference.bib}
\end{document}